\documentclass[1p]{elsarticle}

\usepackage{amssymb}
\usepackage[numbers]{natbib}
\usepackage{rotating}
\usepackage{graphicx}
\usepackage{amsmath}
\usepackage{array}
\usepackage{multirow}
\usepackage{lineno}

\journal{Nuclear Instruments and Methods A}

\begin{document}

\begin{frontmatter}
\title{SHarD: A beam dynamics simulation \\ code for dielectric laser accelerators \\ based on spatial harmonic field expansion}

\author[inst1]{A. Ody}
\author[inst1]{S. Crisp}
\author[inst1]{P. Musumeci}
\author[inst2]{D. Cesar}
\author[inst2]{R. J. England}

\affiliation[inst1]{organization={UCLA Department of Physics and astronomy},
            addressline={475 Portola Plaza}, 
            city={Los Angeles},
            state={CA},
            postcode={90095}, 
            country={U.S.A.}}

\affiliation[inst2]{organization={SLAC National Accelerator Laboratory},
            addressline={2575 Sand Hill Rd}, 
            city={Menlo Park},
            state={CA},
            postcode={94025},
            country={USA}}

\begin{abstract}
In order to demonstrate acceleration of electrons to relativistic scales by an on chip dielectric laser accelerator (DLA), a ponderomotive focusing scheme capable of capturing and transporting electrons through nanometer-scale apertures over extended interaction lengths has been proposed. We present a Matlab-based numerical code (SHarD) utilizing a spatial harmonic expansion of the fields within the dielectric structure to simulate the evolution of the beam phase space distribution in this scheme. The code can be used to optimize key-parameters for the accelerator performance such as the final energy, transverse spot size evolution and total number of electrons accelerated through currently fabricated structures. Eventually, the simulation model will be applied to inform the phase mask profile to be added to a pulse front tilt drive laser pulse using a liquid crystal mask in the experimental setup being assembled at UCLA Pegasus Laboratory.
\end{abstract}


\end{frontmatter}


\section{Introduction}
\label{sec:sample1}
Dielectric Laser Accelerators (DLA) have demonstrated the capability to achieve GV/m accelerating gradients but only over short interaction lengths \cite{High-Field, wootton, Erlangen}. In these experiments, the electron beam is accelerated by a short drive laser pulse which is coupled transversely onto dielectric grating-like structures. In order to obtain high intensities (and therefore high accelerating gradients) while keeping the fluence below the damage threshold, the laser pulse length should be as short as possible. On the other hand, due to the crossing geometry, the interaction length is essentially limited by the laser pulse length to 10-100 $\mu$m. Tilted pulse fronts \cite{hebling} can be implemented in order to extend the interaction length to the mm-scale and achieve higher total energy modulations as demonstrated in \citet{Energy-Gain}.

If even longer interaction lengths are desired, it becomes critical to have better control of the electron beam dynamics in the DLA and provide additional focusing to keep the electrons confined within the sub-micron gap of the acceleration channels over extended travel lengths. This is due to the fact that when electrons sample stable accelerating phases in the DLA, they are also defocused transversely. Conventional magnetic lenses simply cannot provide external focusing fields sufficient to counteract this defocusing force, driving the search for an on-chip focusing solution. 

Various approaches are being explored, but they require complex structure design with fine control of the phase in between various accelerating sections. In particular the proposed Alternating Phase Focusing scheme \cite{APF} exploits the clever introduction of short drift sections to rephase the field seen by the particles and provide focusing both in the transverse and longitudinal phase space. The challenges in this approach are that once the structure is fabricated there are no directly available knobs to fine tune the beam dynamics and adapt the accelerator to the drifting/changing input electron beam and laser distribution. At the same time, tuning the output accelerator parameters such as output energy or phase space distribution is also complicated by the constraints associated to having a pre-built dielectric structure. 

\begin{figure}
    \centering
    \includegraphics[width=\linewidth]{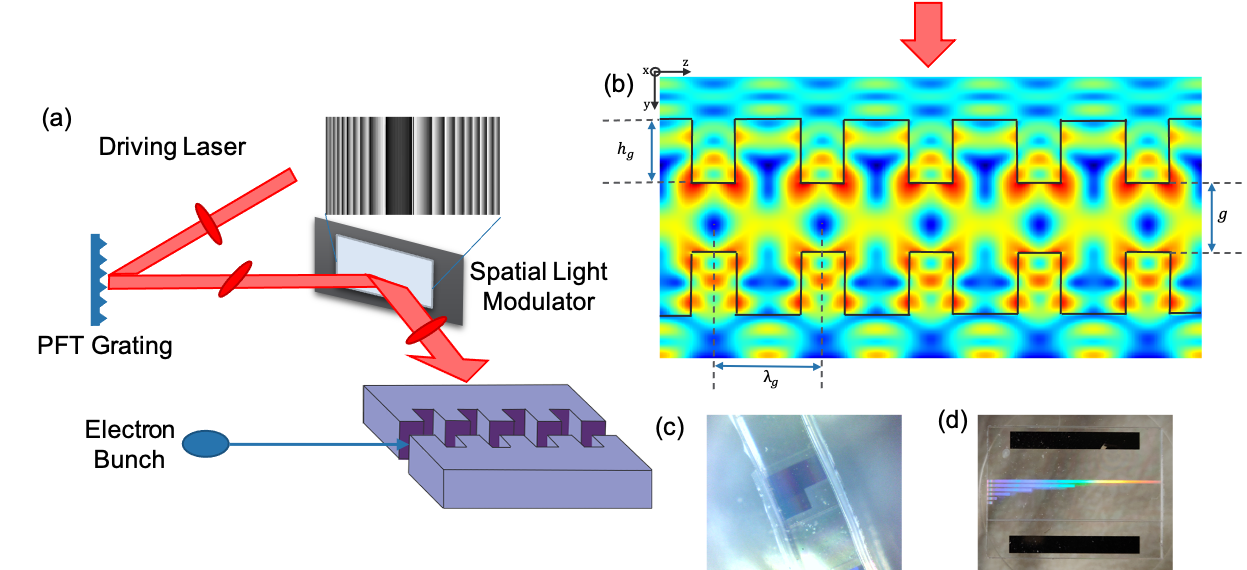}
    \caption{\textbf{Experimental Plans.} The planned experimental setup is diagrammed in \textbf{(a)}. A Lumerical FDTD simulation of the electric field magnitude inside the grating structure, using the 1mm length structure parameters in Table \ref{sim-params}, is shown in  \textbf{(b)}. An example of the 1mm structure \cite{fab:Peralta} and the 2cm structure \cite{YuMiao} are shown in \textbf{(c)} and \textbf{(d)}.}
    \label{fig:diagram}
\end{figure}

A separate approach has been discussed in \cite{cesar:alloptical} where the structure is a simple uniform grating and ``soft" control on the beam dynamics is exerted by imprinting a particular amplitude/phase profile on the drive laser. Pulse front tilt coupling is particularly advantageous here as it allows the use of existing amplitude/phase modulation technologies such as liquid crystal masks to control the field profile seen by the electrons as they travel through the structure. In this case a large amount of freedom is allowed, enabling fine tuning of important parameters such as energy, energy gradient, longitudinal phase space and transverse phase space acceptance. In order to provide transverse containment, Naranjo et al. \cite{Naranjo} advanced the idea of exploiting the ponderomotive (second-order) focusing effect which results from the averaging over fast alternating gradient first-order transverse forces. This concept is not new in accelerator physics and has been applied successfully to control transverse dynamics in proton accelerators \cite{tkalich, masunov, durkin2020}. In practice, in a DLA this is realized by introducing spatial harmonics on the field in the structure so that the differences in the wave phase velocities will cause the electrons to experience a continuous oscillation between focusing and defocusing phases.  

At the UCLA Pegasus laboratory \cite{alesini:Pegasus}, the current experimental goal is to demonstrate the feasibility of a tunable ponderomotive focusing scheme to extend DLA interaction lengths from sub-millimeter out to 1-2~cm and achieve MeV-scale energy gains improving by nearly an order of magnitude over the current state-of-the-art in DLA acceleration \cite{Energy-Gain}. In our setup, detailed in Fig. \ref{fig:diagram} we plan to inject a relativistic electron ($\sim$6MeV) beam from a traditional RF photoinjector source into a double grating structure driven by a few mJ, 780nm, 100fs FWHM laser pulse. To allow for ``soft" control, a sophisticated laser manipulation setup is used to tilt the incoming laser pulse front and ensure that the phase mask imprinted on the beam by a Spatial Light Modulator (SLM) is imaged onto the interaction plane \cite{Sophie}. In the following, we'll discuss the SHarD (Spatial Harmonic DLA) code, the numerical model used to simulate the beam dynamics and fine tune the phase/amplitude mask parameters for these experiments. It is understood that the freedom associated with using an SLM could be taken advantage of in more advanced, ``smart", online optimization schemes where machine learning techniques are applied to feedback on the accelerator output parameters and further improve the performances of the DLA.

\section{FOCUSING BY SPATIAL HARMONICS}

The operating principle of the scheme is to excite different spatial harmonics within the accelerating channel which, due to their different phase velocities, will wash over the beam during the transport and provide a second order (ponderomotive) focusing force on the particles in the accelerator \cite{Reiche:PWT2ndorderfocusing}.

As confirmed using FDTD Lumerical simulations\footnote{www.lumerical.com}, the DLA double-grating structures under study \cite{Peralta} have a relatively low Q-factor and a correspondingly large acceptance bandwidth, much larger than the bandwidth of the input laser pulse. A consequence of this: if we compare the waveform input at the grating layer to the waveform evaluated in the center of the vacuum gap, the input and output pulses are found to be nearly identical, such that we will assume in what follows that the phase and amplitude of the field experienced by the electrons are fully specified by the illumination field.

In this approximation, spatial harmonics can be easily generated simply by adding a sinusoidal phase offset to the drive laser pulse along the direction of propagation. In this case, assuming a monochromatic wave of angular frequency $\omega$ in the DLA, the on-axis complex field amplitude within the structure of periodicity $\lambda_g = 2\pi/ k_g$ can be written as
\begin{equation}
    E_z = -i E_0 e^{ik_g z-\omega t + a \cos \delta_k z}
\end{equation}
where $E_0$ is the on-axis peak field (related to the input field amplitude by the structure factor, typically a number between 0.1 and 0.3 which depends on the details of the double grating such as teeth depth, width and relative alignment) and $a$ and $\delta_k$ are the amplitude and period of the added phase oscillations. We can then use the Jacobi-Anger series expansion
\begin{equation}
    e^{i(k_gz+a\cos(\delta_kz))} = \sum_{m=-\infty}^{\infty}i^m J_m(a)e^{i(k_g+m\delta_k)z}
    \label{Jacobi}
\end{equation}
to write the field as a sequence of spatial harmonics with phase velocities $\beta_m c = \frac{\omega}{k_g+m \delta_k}$. 

In our scheme, discussed in detail in \cite{cesar}, by selecting the proper periodicity $\delta_k$ we tune one of these spatial harmonics (in our case, $r$=-1) to have the same phase velocity as the normalized longitudinal velocity $\beta = \sqrt{1-\frac{1}{\gamma^2}}$ of the externally injected electrons i.e.
\begin{equation}
\frac{\omega}{k_g-\delta_k} = \sqrt{1-\frac{1}{\gamma^2}}
\end{equation}
The resonant harmonic provides a stable acceleration force, while the other harmonics create a rapidly varying field as viewed by the electrons that acts to transversely focus and stabilize the motion. 

Since there should be no phase slippage between the electrons and the wave, the interaction dynamics with the $r$ =-1 harmonic can be described using the concept of resonant phase $\psi_r$. For a constant resonant phase along the DLA, the effective normalized accelerating gradient can be written as $ G = -\frac{\omega}{c} \alpha J_r(a) \sin \psi_r$ where $\alpha = e E_0 / m c \omega$ is the normalized vector potential for the wave. By convention the accelerator has negative resonant phase to accelerate negative charged particles. As the resonant phase approaches $-\pi/2$ the acceleration gradient increases, while the area of the stable bucket in the longitudinal phase space shrinks \cite{wangler}.

In order to keep accelerating electrons in phase with this resonant harmonic, the phase velocity must be adjusted (tapered) along the interaction. Starting from the definition of the wave phase velocity
\begin{equation}
    \beta(z) = \frac{\omega}{k_g-\delta_k+\frac{d\phi}{dz}}
    \label{Eq:resonant_condition}
\end{equation}
and assuming a constant resonant phase along the interaction and a relatively small total energy gain, this tapering is provided by adding a parabolic phase profile to the phase mask with $\frac{d^2\phi}{dz^2} = -(k_g-\delta_k) G / \beta^2 \gamma^3$.

Note that the other harmonics also contribute to the longitudinal phase space evolution, but since they are not resonant they simply induce small amplitude quivering in the beam velocity. To ensure that the acceleration dynamics is not strongly affect by the other spatial harmonics, it is possible to use the Chirikov criterion \cite{Chirikov} which states that in order to avoid chaotic motion the maximum height of the stable bucket in the un-normalized longitudinal phase space $\left(\delta \gamma / \gamma\right)_{max} = \sqrt{4 \beta^3 \gamma \alpha}$, multiplied by a bucket height function which is maximum and equal to 1 when the resonant phase $\psi_r = 0$, needs to smaller than the separation of the accelerating buckets of the other spatial harmonics $\gamma^2\beta^2 \delta_k/ k_g$. 

For an infinitely wide slab-geometry structure where the fields are essentially independent of the x-coordinate, we can write the transverse electric and magnetic field as simple derivatives of the longitudinal field as 
\begin{equation}
E_y = \frac{ih}{\Gamma^2} \frac{\partial}{\partial y} E_z(y,z,t) \quad \text{and} \quad B_x = \frac{i\omega}{c^2 \Gamma^2} \frac{\partial}{\partial y} E_z(y,z,t) 
\end{equation}
where $\Gamma$ is the transverse wavenumber and the dispersion equation for the waves in the channel require $\Gamma^2 + h^2 = \omega^2/c^2$. In our description the field is described by a finite number of longitudinal wavenumbers $h_m = k_g + m \delta_k$ and therefore transverse wavenumbers $\Gamma_m$ depending on the spatial harmonic index $m$.

Assuming a cosh-like transverse dependence for the TM-like accelerating modes in the channel, the transverse force on the electrons can be expressed as
\begin{equation}
    F_y = \Re [\sum_m e E_0 J_m(a) (1-\beta \beta_m) \frac{ih_m}{\Gamma_m} \sinh{(\Gamma_m y)} e^{i(h_m z -\omega t + m \pi/2)}]
\end{equation}
Only one of the terms in the sum is resonant (i.e. has a nearly constant phase), and provides the defocusing force $e E_0 J_r(a) \cos \psi_r$. All the other harmonics, due to the difference of their phase velocities with the electrons, generate an oscillating transverse force. For small y-deviation it is possible to approximate the $\sinh$ with its argument so that the forces are linear. Analyzing the solutions of the y-equation of motion into a slowly varying secular component and a fast oscillation is the basis for the ponderomotive focusing scheme described in \citet{Naranjo}. Note that in practice only one spatial harmonic (other than the resonant wave) would be sufficient to provide the focusing force, but as a byproduct of the sinusoidal phase modulation imparted on the laser, the drive pulse will excite a rich content of spatial harmonics which will all contribute (with a strength depending on their relative phase velocity) to confining the particles in the small gap of the DLA structure.   

\begin{figure}[t]
\begin{tabular}{cc}
  \includegraphics[width=0.48\linewidth]{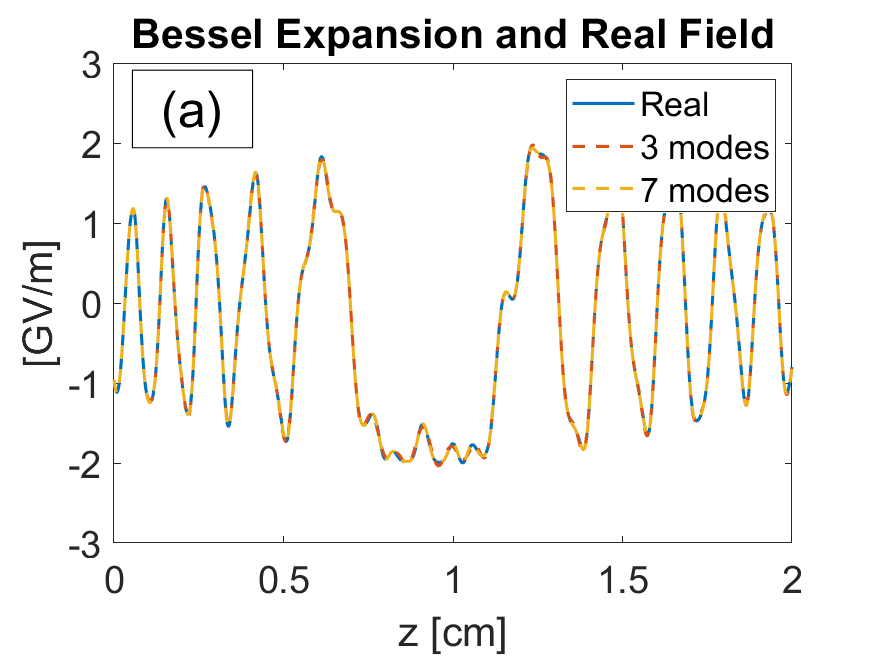} &   \includegraphics[width=0.48\linewidth]{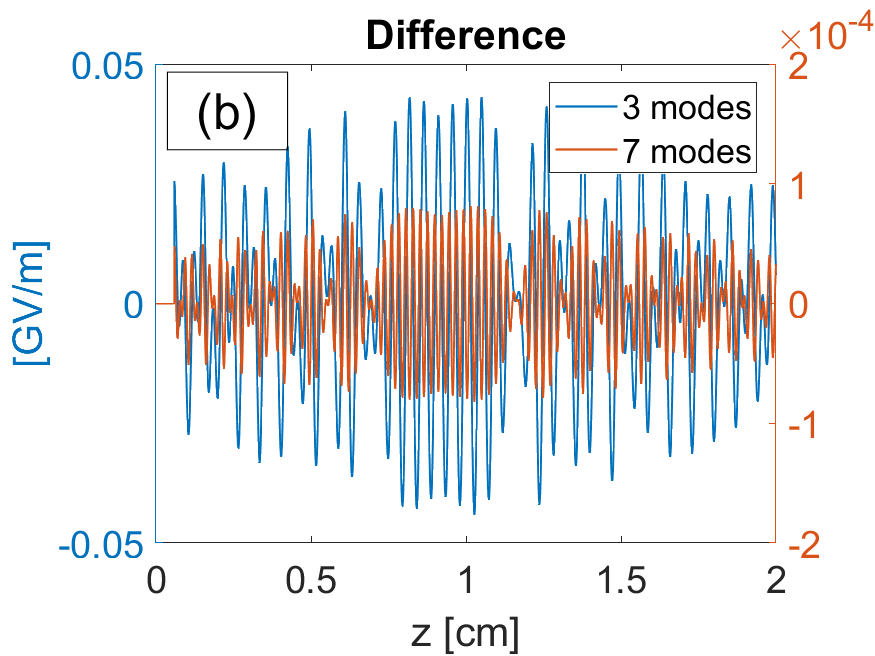} 
\end{tabular}
\caption{\textbf{Truncated Bessel Series Comparison.} \textbf{(a)} Plots of the full exponential form of the electric field and Bessel expansions truncating after 3 harmonics (m=-1 to m=1) and 7 harmonics (m=-3 to m=3). \textbf{(b)} The difference between the actual field and the truncated Bessel expansion approximations shown in a. The first 3 harmonics roughly approximate the beam with 2.5\% maximum difference while the use of higher order terms bring the error to 50 parts in one million, below the uncorrelated energy spread of the electron bunch.}
\label{Fig:expansion}
\end{figure}

In our numerical approach, we will solve for the motion of the electrons under the action of a finite number of finite harmonics. We will leave open the possibility for the modulation $\delta_k$ and the amplitude $a$ of the phase modulation to vary along $z$ as well as the possibility for a slowly varying field amplitude profile $E_0(z)$ along the interaction. Note that in this case the assumption that the fields are fully periodic in $z$ (and hence could be simply written as a sum of spatial harmonics) must be checked as the Jacobi-Anger expansion becomes only an approximation to the actual phase and amplitude profiles when the higher order terms are truncated. For one of the optimized on-axis profiles used in the examples discussed later in the paper, this comparison is done in Fig. \ref{Fig:expansion}. The agreement is excellent and shows that the expansion well reproduced the field experienced by the particles within 100 parts per millions when the first 7 spatial harmonics terms (i.e. m = -(N-1)/2 to m = (N-1)/2) are included in the sum. 
This allows us to efficiently set up the numerical solution to the equations of motion. Further accuracy can be obtained by including higher terms at the cost of simulation time. 


As the scheme is based on the fact that different spatial harmonics have different phase velocities, it is important to retain the term $(1-\beta \beta_m)$ in the transverse force. Note that if the system was perfectly periodic, applying the Panofsky-Wenzel theorem to derive the momentum kicks imparted by the structure would yield a simpler $1/\gamma^2$ dependency. The higher-level approximation was one of the main reasons to develop our own numerical code with respect to the well documented DLATrack6D \cite{DLATrack6D}, which has been developed to simulate dielectric accelerating structures. It is easy to see that the approximation amounts to neglecting $\beta^2 \delta_k/k_g$ with respect to $1/\gamma^2$. In the cases explored numerically below $\gamma^2 \beta^2 \delta_k/k_g \cong 0.1$ and some small difference in the dynamics could be expected. Nevertheless, it turns out that in many cases the approximation is actually extremely well satisfied, especially for non relativistic particles.

\section{MATLAB SIMULATION CODE}
The simulation is done using a Matlab code that tracks the particle coordinates in 6D space. The motion in the transverse x-direction is trivial and fully determined by the initial conditions as the no fields and forces that affect the particle trajectories in that coordinate are negligible in our structures. Initial particle distributions can be input from a different particle tracking code (for example the one used to study the evolution of the beam in the RF photoinjector) or more simply can be randomly generated assuming a given energy, energy spread, pulse length, emittance, spotsize, and number of particles. A prebunched initial distribution can be also used to evaluate the benefits of a bunching stage. The longitudinal coordinates chosen to describe the motion are the resonant spatial harmonic phase $\theta = k_g z - \omega t + r \delta_k z$ and the normalized particle energy $\gamma$. 

The code then solves for the motion of the particles under the forces exerted by the fields in the structure using the Runge-Kutta algorithm. Statistics of the phase space distribution at the exit as well as along the interaction can be used to better understand the accelerator dynamics. A figure of merit for the quality of the acceleration is set by the energy transfer from the laser to the particles determined as a product of the number of particles that are able to propagate at all times within the structure gap and the average energy gain per particle. As this number increases, effects like beam loading and wakefields will come into play, but these are not taken into account in the current version of the code as these effects are not expected to be detrimental for the low bunch charge in planned experiments. 

\newcolumntype{L}{>{\centering}m{2.8cm}}
\begin{table}[t]
    \centering
    \begin{tabular}{L|cccc}
        \hline
        & Structure Length & L & 1mm & 2cm\\
        \hline
         \multirow{3}{*}{\begin{tabular}{c}ELECTRON \\PARAMETERS\end{tabular}} &Initial Total Energy & $m_0c^2 \gamma$ & 6.5 MeV & 6.5 MeV  \\
         & Spotsize & $\sigma_y$ & 0.5 $\mu$m &0.5 $\mu$m \\
         & Normalized Emittance & $\epsilon_{n,y}$ & 0.5 nm-rad& 0.5 nm-rad \\
         \multirow{2}{*}{\begin{tabular}{c}LASER \\PARAMETERS\end{tabular}}& Peak on axis field & $E_0$ & 2 $\frac{GV}{m}$&2 $\frac{GV}{m}$ \\
         & Laser Wavelength & $\lambda$ & 780nm & 780nm \\
         \multirow{3}{*}{\begin{tabular}{c}STRUCTURE \\PARAMETERS\end{tabular}} & Gap Width & g & 800nm & 800nm \\
         & Grating Periodicity & $\lambda_g$ & 800nm & 780nm \\
         & Tooth Height & $h_g$ & 720nm & 700nm/100nm \\
         \multirow{4}{*}{\begin{tabular}{c}OPTIMIZATION \\PARAMETERS\end{tabular}} & Buncher Length & $L_b$ & 100 $\mu$m & - \\
         & Drift Length & $L_d$ & 425 $\mu$m & -\\
         & Phase Periodicity & $\lambda_{\psi}$ & - &$(1200 + 45\frac{z}{L})\lambda$ \\
         & Phase Offset Amplitude & a & - & 0.3rad \\
         \multirow{2}{*}{RESULTS} & Capture Rate & r & $<$14\% & $<$13\% \\
         & Energy Modulation & $\Delta E$ & 1 MeV & 2.5 MeV \\
         \hline
    \end{tabular}
    \caption{SIMULATION PARAMETERS}
    \label{sim-params}
\end{table}The phase and amplitude profile of the field on axis is constructed by a set of user defined inputs that describe the normalized peak wave vector potential $\alpha$, the amplitude profile of the laser, and the amplitude ($a$) and periodicity ($\delta_k$) of the phase oscillations. Both $a$ and $\delta_k$ can be defined as look-up tables and adjusted along the accelerator in an arbitrary way. Tapering of the resonant phase $\psi_r$ is also possible. A $\phi_{tap}(z) = (1/\beta_r -1)kz$ phase profile is also added to ensure that the resonant particle remain in phase with one of the spatial harmonic waves in the structure as it is gaining speed during acceleration. In the experiment, such detailed control would be enabled by overlaying a non-periodic profile on the SLM, which is then relay-imaged onto the DLA structure. 

In the examples presented below, the on-axis field profile is written as 
\begin{equation}
    E(z) = E_0 e^{\frac{-(z-L/2)^2}{2\sigma^2}}e^{i (k_g z - \omega t + a \sin \delta_k z + \phi_{tap}(z))}
\label{eq:phase}
\end{equation}
where $L$ is the length of the DLA structure and a gaussian profile of rms size $\sigma$ centered half-way in the structure is used to describe the field amplitude.

Essentially, we will be calculating the interaction of the particles with a finite number of spatial harmonics modes with amplitude extracted by this field. Including only a finite number of Jacobi-Anger expansion terms the particle positions and momenta are updated using the Runge-Kutta method with the following derivatives
\begin{equation}
    \begin{split}
        \frac{d\gamma}{dz} &= \sum_m [k -\alpha J_{m}(a) \sin \left(\theta + (m-r)(\delta_k z + \frac{\pi}{2})\right)]\\
        \frac{dy'}{dz} &= \sum_m \frac{k \alpha J_m(a)}{\gamma} h_m y  \left(1-\beta_m \frac{\sqrt{\gamma^2-1}}{\gamma}\right)  \cos \left(\theta + (m-r)(\delta_k z + \frac{\pi}{2})\right)\\
        \frac{d\theta}{dz} &= k\left(1  - \frac{\gamma}{\sqrt{\gamma^2-1}} + \frac{1}{k}\frac{d\phi_{tap}}{dz}\right)\\
        \frac{dy}{dz} &= y'\\
    \end{split}
\end{equation}
where $k = \omega/c = 2\pi / \lambda$ is the laser wave vacuum wavenumber, $s$ is the coordinate along the grating structure and the slowly-varying phase $\phi_{tap}(z)$ has been added to $\theta$, so that the resonant particle phase trajectory along the structure is simply constant i.e. $\theta = \psi_r$.

The code runs fast on a single CPU machine and can integrate the equation of motion for 100k particles in 100 s going over 1000 grating periods. Optimization of the step size is performed to reduce the simulation time while at the same time minimizing the error due to the non-symplectic integration on the particle trajectories. Varying the step size between 0.1 grating periods to 10 grating periods saw no change to the number of particles focused through the structure, although larger step sizes led to a divergence in the results. Outputs include the particle phase space distributions at each point, resonant energy gain, and the total number of accelerated and captured particles.

\section{SIMULATION CASES}

\begin{figure}[t]
\begin{tabular}{ccc}
   \multirow{2}{*}[0.55in]{ \includegraphics[width=0.45\linewidth]{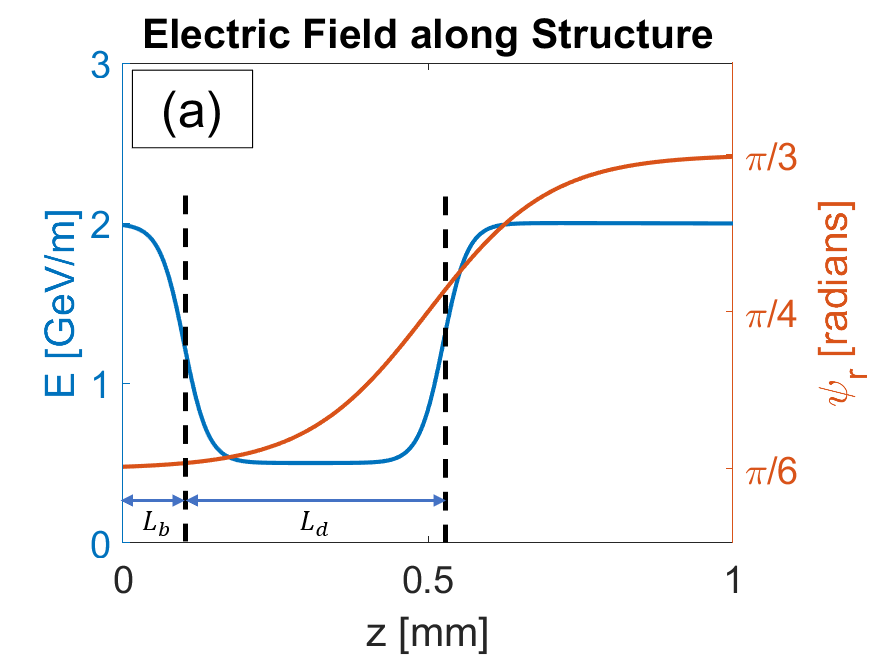}}&\includegraphics[width=0.2\linewidth]{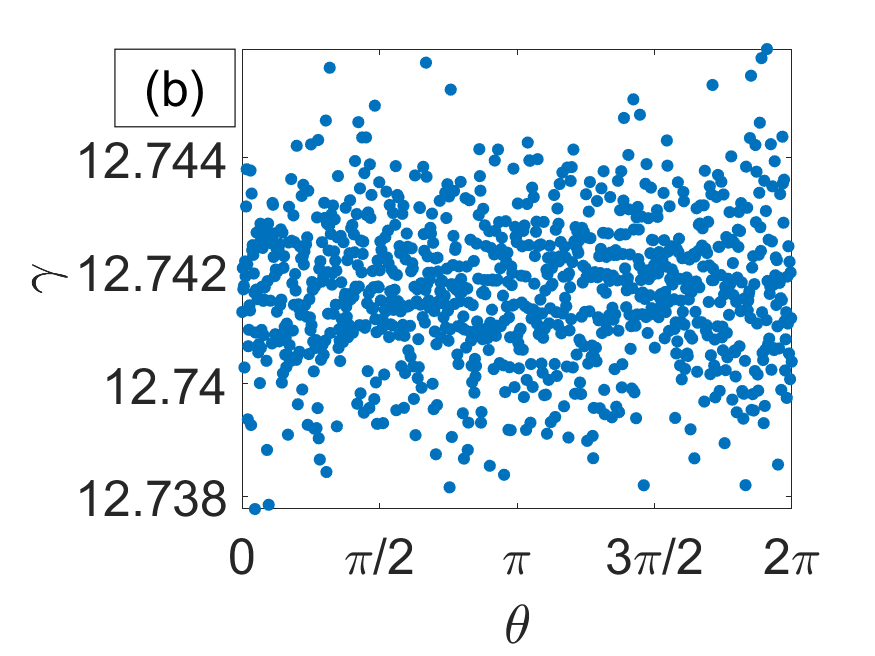}  &  \includegraphics[width=0.2\linewidth]{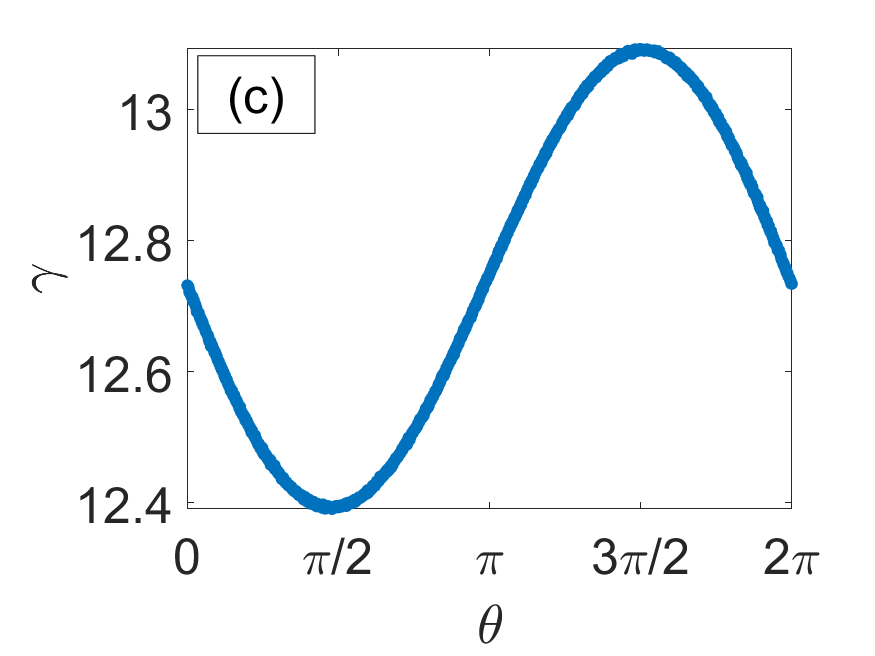}  \\
   & \includegraphics[width=0.2\linewidth]{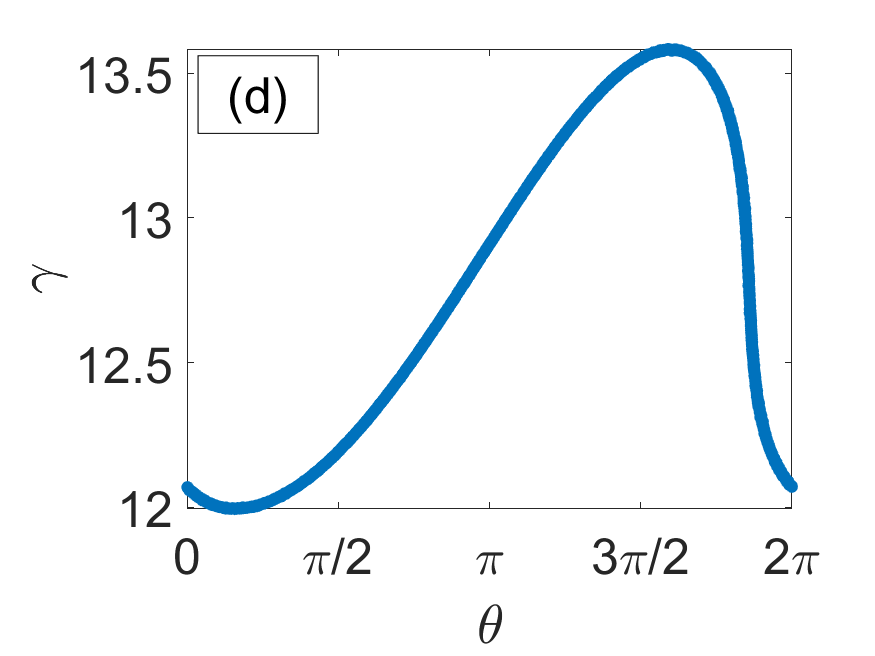} &  \includegraphics[width=0.2\linewidth]{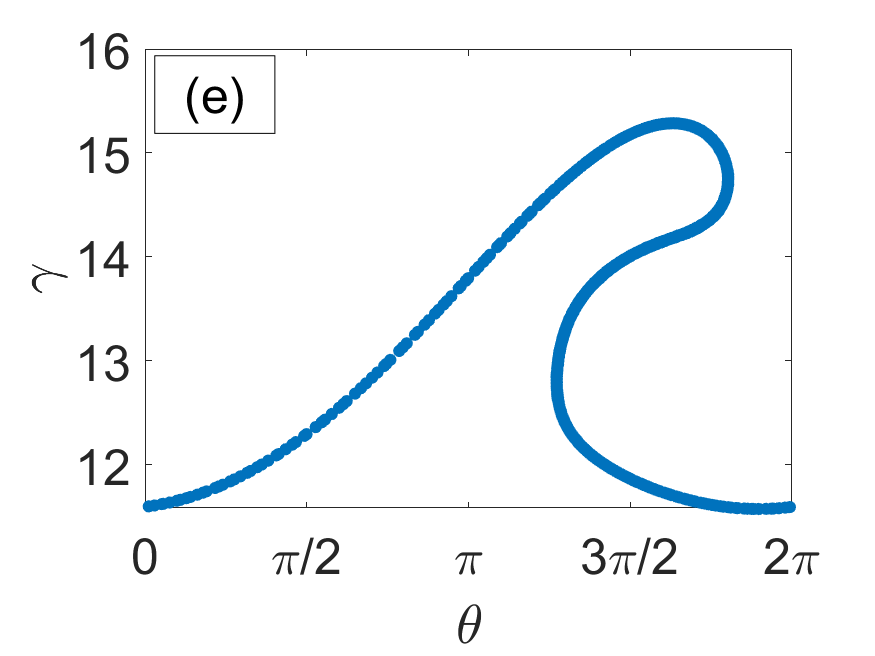} \\
\end{tabular}
\caption{\textbf{1mm Structure Simulation.} \textbf{(a)} Field strength and phase along the structure. The laser intensity can be modulated by shaping along the relay-imaged coordinate the cylindrical focusing lens strength applied to the driving laser pulse using the Spatial Light Modulator. \textbf{(b)} shows the initial energy distribution at z=0mm, \textbf{(c)} at the first vertical line z=0.1mm, just after the inital buncher, \textbf{(d)} at the second vertical line z=0.525mm, after the drift section, and \textbf{(e)} at z=1mm, the final energy distribution out of the structure}
\label{Fig:1mm}
\end{figure}

We focus now on two simulation cases, exploring different capabilities of our proposed scheme using the SLM for two different DLA structures. The first case uses an amplitude modulation of the driving laser to separate in a relatively short structure (1mm) a short bunching section and an energy-boost section separated by a drift section where the field amplitude is intentionally maintained very low. The second case explores using the SLM to impart a sinusoidal phase modulation to excite spatial harmonics within the structure and ponderomotively focus the electrons within the sub-micron channel of the structure over an interaction length of 2 cm. Both of these structures are homogeneous double-grating structures with parameters listed in Table \ref{sim-params} and have been fabricated at the Stanford nanofabrication center \cite{fab:Peralta,YuMiao} and will be available for the experimental tests planned at the UCLA Pegasus Laboratory.

\subsection{Amplitude Modulation - 1mm Structure Length}
In a first test we plan to use the structures previously utilized for the majority of all relativistic DLA experiments up to date \cite{wootton, Peralta, cesar}. These double grating structures have a total length of 1 mm and a periodicity of $\lambda_g = 2 \pi/ k_g =$800nm.

Here we can take advantage of the SLM configuration and the cylindrical-lens-based optical system which only images the SLM plane onto the DLA accelerator in one plane to shape the intensity profile of the laser along the beam propagation direction by imparting different parabolic phase profiles (i.e. focusing strength) along the non-imaging dimension. This essentially results in regions of higher and lower intensity on the DLA. In this way, it will be possible to modulate the electric field strength seen by the electrons and create within the same structure effectively a buncher and drift section optically, without needing to design the structure specifically ahead of time. With an on axis peak field of $E_0$ = 2 GV/m driving field the energy gain after a short section of DLA of length $L_1$ will be $\Delta \gamma = \frac{e E_0 L_1}{m_0c^2} \sin k z$. Moderately relativistic electrons experience a relatively large $R_{56} = L_2/\gamma^2$ when propagating in a drift. The energy modulation gets converted into maximum bunching when $R_{56} \frac{eE_0L_1}{m_0c^2 k} = 1$. We can then solve for the lengths $L_1$ and $L_2$ of high and low intensity (bunching and drift) region in the DLA subject to the constraint of the real estate available i.e. $L_1 + L_2$ = 0.5 mm so that the last half of the structure can be used for acceleration. The analytical first order solution yields $L_1$ = 170 $\mu$m and $L_2$ = 330 $\mu$m, close to the results of the numerical optimization. In practice, the lengths of the bunching section and following drift can be adjusted and optimized online. Additionally, the phase for each section can be controlled separately by use of the SLM, allowing the resonant phase of the accelerating section to be ramped once the particles have been captured. Tuning the exit phase (set to $\pi/3$ in our simulations) will allow to tune the final beam energy simply adjusting the phase profile on the SLM. In Fig. \ref{Fig:1mm}a we show the electric field strength and phase along the structure. Figures \ref{Fig:1mm}b-e show the energy distributions of the beam at key points along the structure.

\subsection{Phase Modulation - 2cm Structure Length}
\begin{figure}[t]
\begin{tabular}{cc}
  \includegraphics[width=0.45\linewidth]{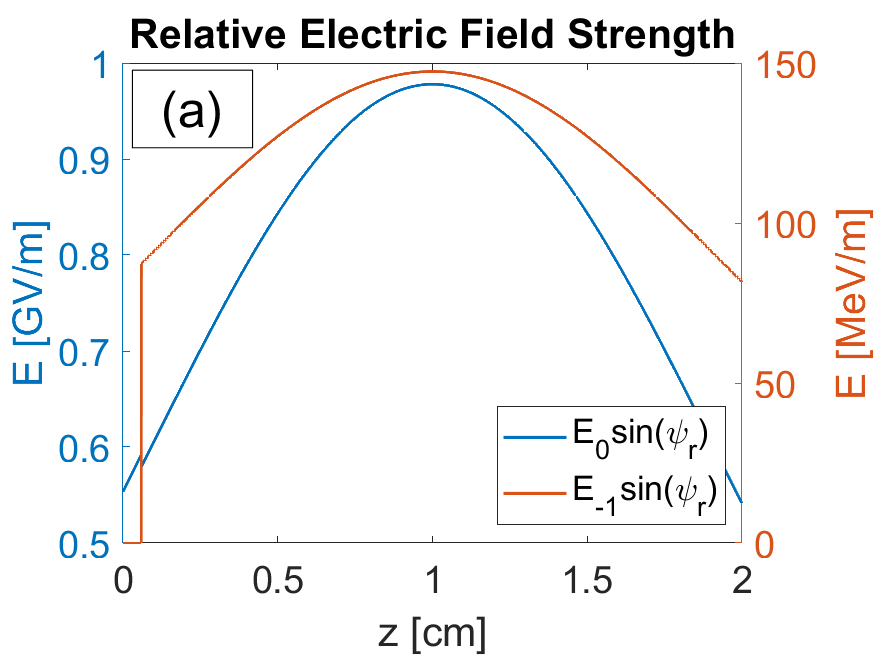} &   \includegraphics[width=0.45\linewidth]{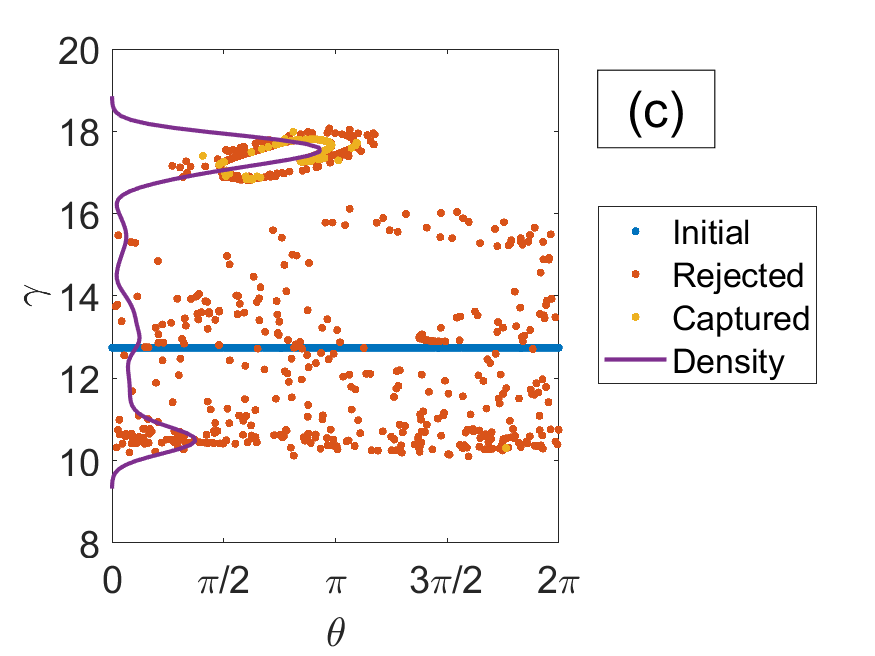} \\
 \includegraphics[width=0.45\linewidth]{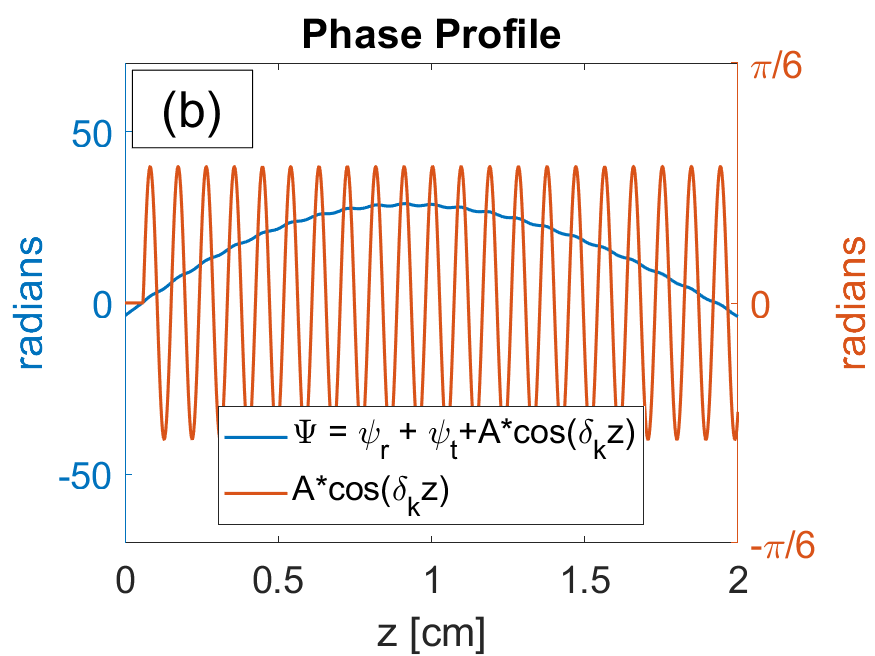} &   \includegraphics[width=0.45\linewidth]{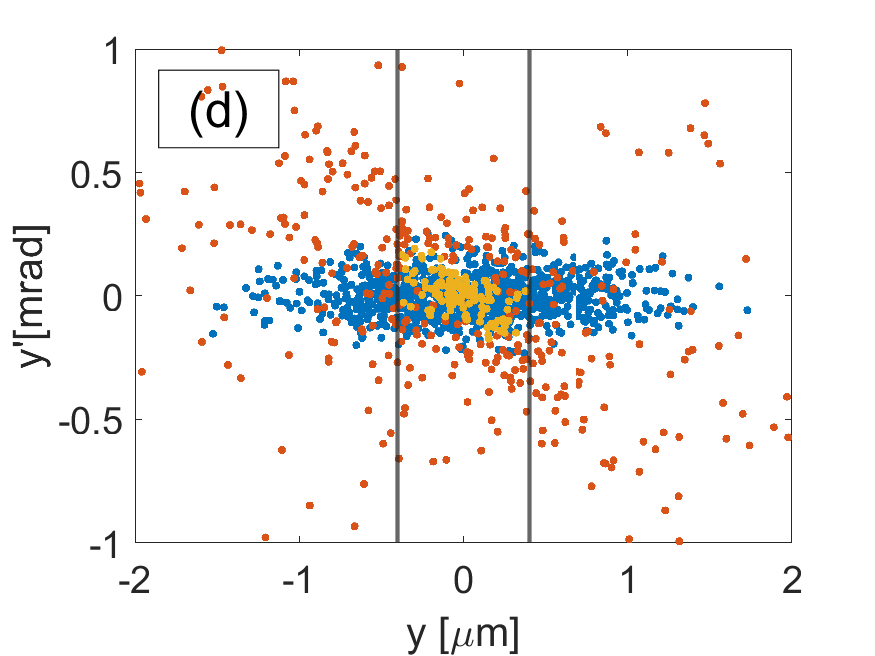} \\
\end{tabular}
\caption{\textbf{2cm Structure Simulation.} \textbf{(a)} Electric field strength of the 0th order and resonant -1 order modes. \textbf{(b)} Phase profile of the driving laser, including parabolic taper to keep resonance and sinusoidal modulation to excite the spatial harmonics or sidebands. \textbf{(c)} Output energy spectrum of the beam with initial spotize 0.5$\mu$m and emittance 0.5nm-rad. Particles both accelerated and focused within the channel are colored in yellow, and account for 13\% of the total particles. \textbf{(d)} Final phase space of the beam, with vertical lines marking the size of the channel}
\label{2cm-fig}
\end{figure}

For longer structures, it is not enough to just control the intensity of the laser along the structure. Rather, phase control is necessary to focus any particles through the channel gap. We focus our simulation efforts here on the 2cm maximum structure length we are planning to employ in our experiment.  The phase profile of the driving laser, plotted in Fig. \ref{2cm-fig}b has three components as shown in Eq. \ref{eq:phase}. The first one is simply the resonant phase which in this simulation is kept constant along the beamline, but could be ramped in a further optimization stage similarly to the 1~mm structure once the particles are first captured. The second component is the tapering phase which is added to keep the resonant spatial harmonic in phase with the accelerating electrons. This is directly dependent on the average gradient in the accelerator and it is a function of the available laser gradient and on the desired ratio of the power in the different spatial harmonics (i.e. the amplitude of the phase modulation) along the structure. The final component is the sinusoidal oscillation which provides the ponderomotive focusing, which is free to be defined by the parameters shown in Table \ref{sim-params}. To make the simulation more realistic we assumed a gaussian profile for the drive laser pulse with a 1.2 cm 1/$e^2$ transverse spot size (which thanks to the pulse-front-tilt geometry is seen by the electrons as the longitudinal profile of the field). This can be adjusted using a telescope on the incoming laser pulse transport line to adjust its dimensions. The design value is chosen to effectively cover the entire structure length and at the same time the entire SLM chip. To maximize the longitudinal acceptance, a short section in the first mm of the DLA is driven by an unmodulated laser. The length of this section is tuned and optimized to maximize loading in the stable region of phase space, so that the particles arrive to the region with the phase modulation already prebunched. A simplex based optimization is performed to tune the spatial harmonics within the structure and maximize the energy transfer of the laser to captured particles as discussed above. The result of this optimization is that in this scheme, most of the laser energy (over $90 \%$) is coupled into the fundamental E$_0$ mode, which is used for focusing, while the next strongest mode, the E$_{-1}$ mode, which is used to accelerate the particles, has only $\sim 2\%$ of the power, resulting in a steep decrease in the average gradient. The relative field strengths are shown in Fig. \ref{2cm-fig}a and the output energy spread and phase space of the electrons are shown in Fig\ref{2cm-fig}c-d. For an electron beam with initial rms transverse spot-size 0.5$\mu$m and emittance 0.5nm-rad, about 13\% of particles are both transported through the 800nm channel without hitting the grating walls and are accelerated by 2.4 MeV on average. 

Further study is needed to investigate a potential middle ground, that targets a higher resonant energy gain, while satisfying a threshold capture rate of particles. Alternatively, the accelerating spatial harmonic could be increased in energy along the structure and the focusing decreased once the particles have been successfully captured into the accelerating bucket. The overall goal would be to decrease the amount of structure length required to accelerate a certain percent of particles.

\section{CONCLUSION}
This paper presents a Matlab simulation code which has proven to be a useful tool for simulating the phase masks' effects on the electron beam in double grating DLAs.
While none of the concepts presented are particularly new, what we present here is a novel application of these ideas and packaging of the various concepts into a numerical tool which can run fairly quickly on a single processor and be used to optimize the dynamics and to provide feedback for tuning the SLM mask. Moving forward, the simulation tool can be further improved by adding an interpreter so that phase masks that would be sent to the SLM can be used as inputs directly, rather than generated from parameters, and utilizing machine learning methods to improve parameter search and optimization. The flexibility offered by the choice to develop the calculation using a Matlab interface will eventually allow this code to be embedded in a control loop system for the actual experiment, such that the phase mask can be changed to optimize the output of the DLA directly. This technique can also be applied to future DLA sources and low energy accelerating sections to create beamlines capable of reaching high electron energies on a chip using soft-control on the accelerator parameters.

\section{ACKNOWLEDGMENTS}
This work has been supported by the ACHIP grant from the Gordon and Betty Moore Foundation (GBMF4744) and by U.S. Department of Energy grant DE-AC02-76SF00515. This material is based upon work supported by the National Science Foundation Graduate Research Fellowship Program under Grant No. DGE-1650604. Any opinions, findings, and conclusions or recommendations expressed in this material are those of the author(s) and do not necessarily reflect the views of the National Science Foundation.

\bibliographystyle{els-article} 
\bibliography{SHarD}

\end{document}